\documentclass[fleqn,twoside]{article}
\usepackage{espcrc2}

\usepackage{graphicx}
\usepackage[figuresright]{rotating}

\newcommand{\ignore}[1]{}
\newcommand{\be}{\begin{equation}}
\newcommand{\ee}{\end{equation}}

\newcommand{\half}{ {\scriptstyle \frac{1}{2} } }

\newcommand{\AmS}{{\protect\the\textfont2
  A\kern-.1667em\lower.5ex\hbox{M}\kern-.125emS}}

\title{Is the graviton a domain wall glueball?}

\author{Richard C. Brower\address{Physics Department\\
        590 Commonwealth Avenue \\
        Boston University\\
        Boston MA 02215}\thanks{
This work was supported in part by the Department of Energy under
Contracts No. DE-FG02-91ER40676}}       
\begin{document}

\begin{abstract}
Strong coupling calculations for the glueball spectrum in an $AdS^7$
black hole are modified by introducing an UV cut-off at a Planck
brane. A new normalizable state in the tensor spectrum is found which
reproduces Einstein-Hilbert gravity up to exponentially small corrections due to off shell
mixing with massive glueballs. 
\vspace{1pc}
\end{abstract}

\maketitle

\section{INTRODUCTION}

Recent developments in string theory have changed our perspective on
the ancient puzzle on how to construct the QCD string.  The conventional
wisdom had been that the fundamental superstring for gravity operates
exclusively at or near the the Planck scale with the extra 6 dimension
compactified in Planckian units. The standard model (and therefore
QCD) would appear as simply a low energy effective field theory long
after all stringy effects have disappeared. In this scenario the fact
that confinement of chromodynamic flux and the large N expansion might
also exhibit stringy features is essentially accidental, with no
physical relation to the strings of quantum gravity.

With the discovery of Maldacena's string/gauge
duality~\cite{maldacena}, we now have explicit examples of the
equivalence between 10-d superstring and 4-d Yang-Mills theory. The key
new ingredient is a strongly ``warped'' radial (or 5-th) dimension as
illustrated in the weak/strong duality between strings in $AdS^5
\times S^5$ and 4-d ${\cal N} = 4$ super Yang Mills.  The redshift in
the warped axis interpolates between low energy (IR) and high energy
(UV) physics.  Extensions of the string/gauge duality have been
suggested by Witten~\cite{wittenT} and others for non-SUSY QCD like
theories exhibiting discrete glueball spectra, confinement, etc.  In
addition Randall and Sundram have shown how brane world gravity maybe
realized in AdS space with the Planck/TeV hierarchy separated
exponentially as function of the proper distance in the warped axis.

Here I review our recent calculations of the glueball spectrum at
strong coupling~\cite{bst4d} and its extension to include brane world
gravity~\cite{bstAdSbh}. The goal is to develop a toy model which is
illustrative of this new approach to the QCD string.

\section{GLUEBALLS IN ADS BLACK HOLE}

The $QCD_4$ model proposed by Witten~\cite{wittenT} modifies
the 11-d $AdS^7 \times S^4$ metric solution to M-theory
or its low energy limit 11-d supergravity,
\begin{eqnarray}
 S &=& \frac{1}{2 \kappa_{11}} \int d^{11}x \sqrt{-g_{11}} \; ( R_{11} -  |F_4|^2) \nonumber \\ 
&-& \frac{1}{12 \kappa_{11} }\int A_3 \wedge F_4 \wedge F_4 + \mbox{fermions}  \; ,
\end{eqnarray}
by compactified both the 11-th axis reducing it to IIA string theory
and a second ``thermal'' (or $\tau$) axis with anti-periodic Fermionic
boundary conditions to break the conformal and super symmetries.  The
resultant background metric is an $AdS^7 \times S^4$ black hole,
$$
ds^2 = \frac{1}{r^2 - r^{1-d}}\; dr^2  
+ (r^2 - r^{1-d})\;d\tau^2 +   r^2 \eta_{\mu\nu} dx^\mu  dx^\nu  
$$
The minimum value $r > r_{min}=1$ is a co-ordinate singularity at the
black hole horizon.
\begin{figure*}
\includegraphics*[width =60mm , height = 80mm]{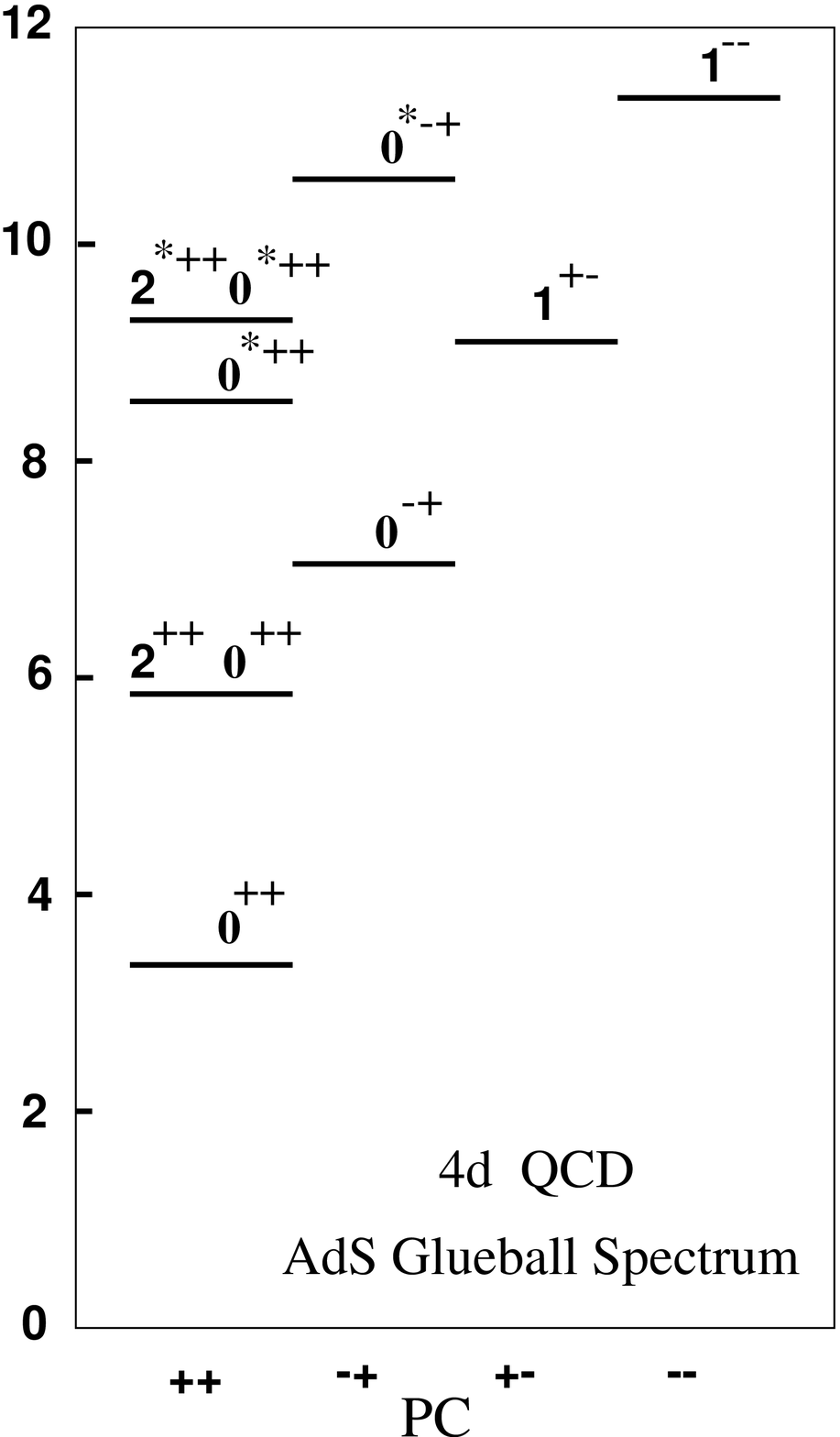} \hskip 2 cm
\includegraphics*[width =75mm , height = 80mm]{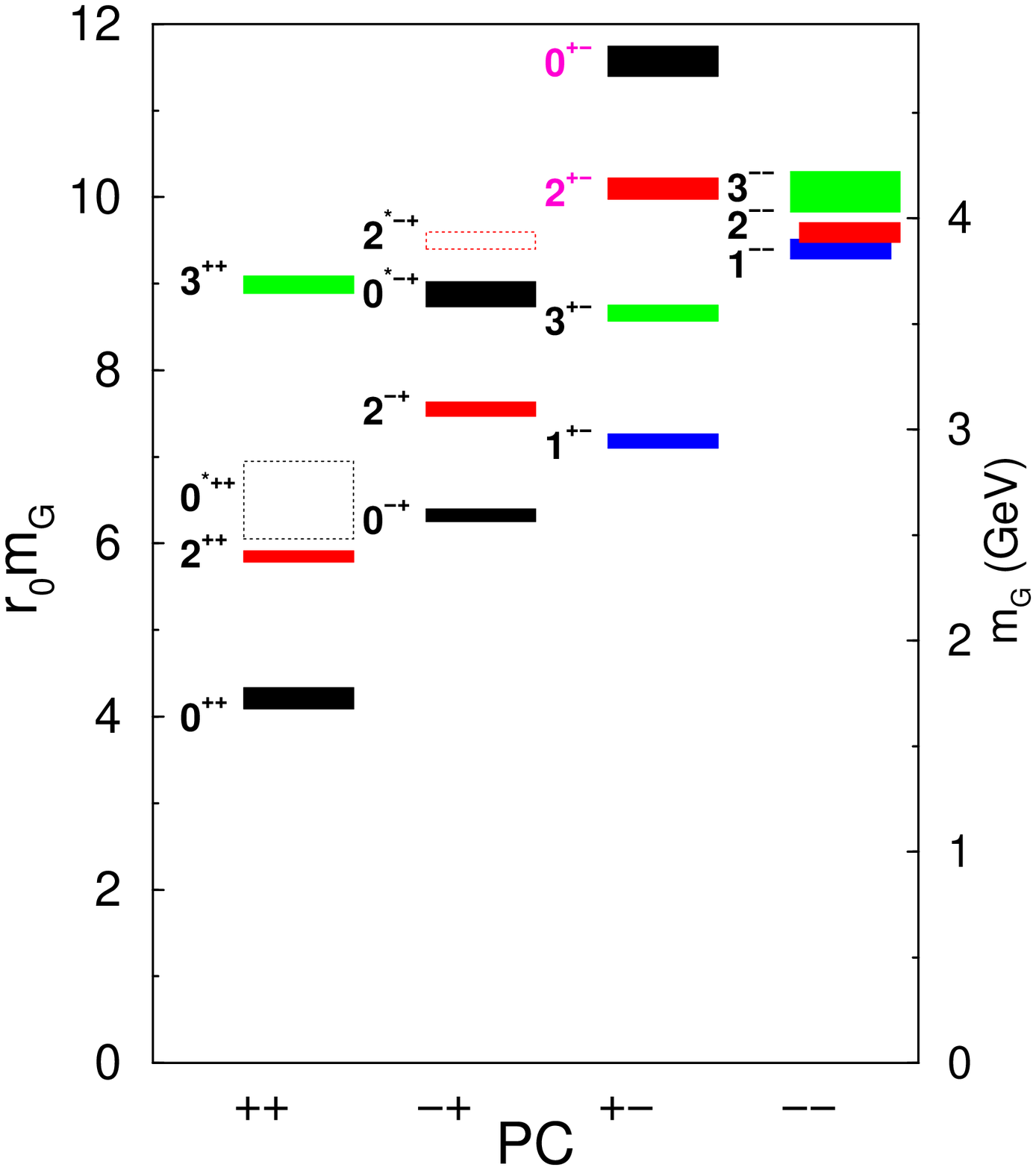}
%
\vskip - 0.5 cm
\caption{Comparison of AdS glueball spectrum for $QCD_4$ in strong coupling
(left) and the Morningstar \& Peardon lattice spectrum (right). The
scales are adjusted to fit $2^{++}$ with $1/r_0 = 410$ Mev. }
\label{fig:comparison}
\end{figure*}

In the strong coupling 'tHooft limit ($g^2 N \rightarrow \infty$), the
inverse QCD string tension vanishes like $\alpha' \sim
(R_{ads}/r_{min})^3 \alpha'_{string} \sim 1/(g^2 N \Lambda^2)$, where
$\Lambda$ sets the scale of the glueball spectrum.  Hence the string
collapses to a point so the glueball spectrum can be found as the
normal modes in the supergravity fields $G_{MN},A_{MNL}$ (see Table
below). 

\smallskip
\begin{center}
\begin{tabular}{|c|c|c|c|}
%
%
%
\hline
$\;\;\; G_{\mu\nu}\;\;\; $   & $G_{\mu,11} $       & $G_{11,11}$  & $m_0$   
\\
\hline
$G_{ij}$              & $C_i$      & $\phi$   &                \\
$2^{++}$              &  $1_{(-)}^{++}$     & $0^{++}$    & 4.7007 \\
\hline
$G_{i\tau}$           & $C_{\tau}$   &              &                   \\
$1_{(-)}^{-+}$              & $0^{-+}$      &               &  5.6555  \\
\hline
$G_{\tau\tau}$       &               &               &                 \\
$0^{++}$      &                       &                 & 2.7034    \\
\hline
\hline
$A_{\mu\nu,11}$         & $A_{\mu\nu\rho}$ &&  $m_0$ (Eq.)                  \\
\hline
$B_{ij}$   & $C_{123}$     &&    \\
$1^{+-}$               & $0_{(-)}^{+-}$      &&   7.3059  \\
\hline
$B_{i\tau }$   & $C_{ij\tau}$  &&         \\
$1_{(-)}^{--}$               & $1^{--}$      &&     9.1129            \\
\hline
\end{tabular}
\end{center}
\smallskip

The entire strong coupling glueball spectrum that lies in
the superselection sector of $QCD_4$, has been
calculated~\cite{bst4d} and its comparison with the numerical values
determined in lattice QCD (Fig.~\ref{fig:comparison}) reveals
surprisingly good agreement in view of the extreme approximation of the
strong coupling limit.

%
\setlength{\unitlength}{0.6 mm}
\begin{picture}(100,65)
\linethickness{.65mm}
\put(10,35){\vector(1,0){80}}
\put(20,30){\line(0,1){10}}
\put(75,30){\line(0,1){10}}
\put(22,40){$r = 1$ (IR)}
\put(85,40){$r = r_c$ (UV)}
\put(10,30){$0$}
\put(85,30){$r \rightarrow \infty$}
\qbezier(20,35)(20,65)(75,65)
\qbezier(20,35)(20,5)(75,5)
\qbezier(75,65)(70,35)(75,5)
\put(80,60){$\tau$}
\qbezier(75,65)(80,35)(75,5)
\end{picture}

All the lowest $J^{PC}$ states are present, requiring contributions
from all the bosonic fields in IIA supergravity.  Even the radial
excitations (illustrated by $0^{*-+}$ in Fig.~\ref{fig:comparison})
are the right order of magnitude.  The absence of higher spins, such
as $3^{+\pm}$, is a direct consequence of the strong coupling (or
infinite tension) limit whereby all ``orbital'' modes are sent to
infinity.

\section{BRANE WORLD GRAVITY IN ADS BLACK HOLE}

Next we have introduced a Planck brane as a UV cut-off at $r = r_c$ by
reflecting the metric ($r \rightarrow r_c/r^2$) at a $Z_2$ orbifold
and adding a positive tension brane world action~\cite{bstAdSbh}. The
essential new feature is that in addition to massive tensor glueballs,
there is now a new normalizable tensor eigenfunction with zero mass --- the
graviton.  After scaling out the warp factor $g_{\mu \nu} \simeq r^2 (
\eta_{\mu \nu} + h_{\mu\nu})$ the transverse tensor amplitude
$h^{\perp}_{\mu\nu}= \epsilon_{\mu\nu}(p) T(r) e^{i p x}$, obeys the
same equation as a minimally coupled scalar.

For comparison with the Randall-Sundram solution, we convert to the
proper distance $y(r)$ from the horizon, $ y(r) = \int^r_b
dr/\sqrt{f(r)}$, and introduce an integrating factor,$\Psi =  e^{- W(y)} T$,
so that the tensor equation has the SUSY form,
\be
[\frac{\partial}{\partial y} - W'(y)][ \frac{\partial}{\partial y} + W'(y)]
\Psi(y)  = \frac{p^2}{r^2} \Psi(r)
\ee
with prepotential $W(y) = - \half \log[\;\sinh((d+1) k (|y_c - y| -
y_c))\;]$. The effective potential $V_{eff}(y) = (W')^2 - W''$ is plotted for
$y \in [0, y_c]$ in the figure below.

\smallskip
\includegraphics*[width =70mm , height = 65mm]{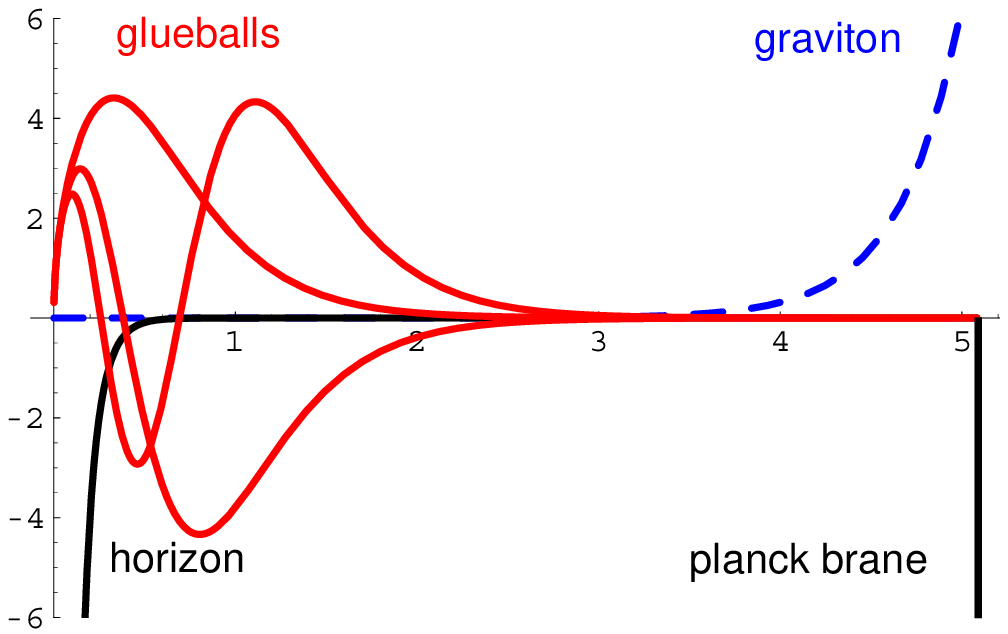}
\smallskip

Note the ``volcano'' potential at the Planck brane is a delta function
at the edge of the disk ($y_c = y(r_c)$ for all $\tau$) with a single
zero mass bound state (dashed line) , the graviton wave function: $\Psi
= \Psi_0 \exp(- W(y))$.  The black hole horizon at the origin ($y = 0$
or $r = r_{min}$) is represented as a double pole in the potential
$W(y)$ which localizes the massive glueball wave functions (solid
lines). The mass hierarchy between the glueballs and the Planck mass
is a reflection of the fact that glueball wave functions are
concentrated near the black hole horizon where the graviton
wave function is exponentially suppressed.

\section{REMARKS}

In addition to the graviton, careful analysis reveals a radion mode
(flexing the proper distance separating the Planck brane from the
black hole horizon) and a massless vector. At low energies the Planck
brane action can be modified, consistent with covariant
energy-momentum conservation, lifting these to small positive
masses. If we were working in $AdS^6$, these three would be the only
``domain wall'' states added to the glueball spectrum.  However the
boundary of $AdS^7$ is $R^{3,1} \times S^1 \times S^1$ with an extra flat
compact $S^1$ along the 11th axis. Hence this analysis has in fact led
us to 5-d gravity with standard Kaluza Klein (KK) compactification and
in addition to the graviton the usual KK gravi-photon and dilaton.
At
strong coupling there is an accidental $O(4)$ symmetry, causing for
example the $2^{++}/0^{++}$ degeneracy seen in the spectrum (see
Fig.~\ref{fig:comparison}).  However this degeneracy will be broken at
weak coupling since the 11-th axes is differentiate from the spatial
axes as the $S^1$ on which the M-theory membrane is wrapped to get IIA
strings. So the mechanism that lifts this symmetry must evidently give
mass to the KK photon/dilaton as well.

\end{document}